# Localized Charge Transfer Process and Surface Band Bending in Methane Sensing by GaN Nanowires

*Avinash Patsha,*[‡,*] *Prasana Sahoo,*[†] *S. Amirthapandian,*[#] *Arun K. Prasad,*[‡] *A. Das,*[‡] *A. K. Tyagi,*[‡] *Monica A. Cotta,*[†] *Sandip Dhara*[‡,*]

[‡] Surface and Nanoscience Division, Indira Gandhi Centre for Atomic Research, Kalpakkam-603102, India

[†] Instituto de Física "Gleb Wataghin", Universidade Estadual de Campinas, 13083-859 Campinas SP, Brazil

[#] Metal Physics Division, Indira Gandhi Centre for Atomic Research, Kalpakkam-603102, India

**Corresponding Authors**

*E-mail: avinash.phy@gmail.com.

*E-mail: dhara@igcar.gov.in.

*Abstract*

The physicochemical processes at the surfaces of semiconductor nanostructures involved in electrochemical and sensing devices are strongly influenced by the presence of intrinsic or extrinsic defects. To reveal the surface controlled sensing mechanism, intentional lattice oxygen defects are created on the surfaces of GaN nanowires for the elucidation of charge transfer process in methane ($CH_4$) sensing. Experimental and simulation results of electron energy loss spectroscopy (EELS) studies on oxygen rich GaN nanowires confirmed the possible presence of 2($O_N$) and $V_{Ga}$–3$O_N$ defect complexes. A global resistive response for sensor devices of ensemble nanowires and a localized charge transfer process in single GaN nanowires are studied *in situ* scanning by Kelvin probe microscopy (SKPM). A localized charge transfer process, involving the $V_{Ga}$–3$O_N$ defect complex on nanowire surface is attributed in controlling the global gas sensing behavior of the oxygen rich ensemble GaN nanowires.

**Keywords** : SKPM, surface band bending, methane sensing, GaN, EELS

# 1. Introduction

Understanding the physical and chemical processes at the surfaces of semiconductor nanostructures is of paramount importance for developing nanostructures based energy and sensor devices. As an example, efficient $H_2$ evolution in photoelectrochemical water splitting has been realized for energy applications by controlling the surface Fermi level pinning with different kinds of dopants in GaN and InGaN nanowires.[1-4] Functionalization of semiconductor nanostructure surfaces with different metal clusters and functional groups resulted in realization of highly sensitive and selective gas and biochemical sensors.[5-7] The underlying physicochemical processes involved in the aforementioned device operations are absorption of photons or adsorption of gaseous molecules, and subsequent transfer and transport of charge on semiconductor surfaces.[8] However, owing to the high surface-to-volume ratio at nanoscale, the semiconductor nanostructure surface usually shows complex behaviours towards the above three fundamental processes, and hence alters the device performances unambiguously.

For chemisorbed species during the sensing, the charge transfer on the surface is driven by the active sites available on the semiconductor surface.[8] The active sites are governed by intrinsic and extrinsic defects in semiconductors, metal-semiconductor Schottky junctions or some functional groups attached to the surface. In the case of oxide semiconductor surfaces ($SnO_2$, ZnO, and $TiO_2$ and others), sensing takes place via chemisorbed oxygenated species ($O_2^-$, $O^{2-}$) serving as active sites for charge transfer and enhanced sensitivity.[9,10] Similarly, the functionalized metal nanoparticles like Ag, Au, Pt and Pd create active sites for the selective or low temperature sensing.[8,11-14] Whereas, thermally stable and chemically inert pure group III-nitride semiconducting surfaces (such as GaN and AlGaN) utilize the Schottky junction with metal or heterojunctions with other semiconducting surfaces as active sites in the sensing process.[5,15-21] Despite the potential applications of group III-nitride nanowires as LEDs, lasers, high electron mobility transistors (HEMTs), logic gates, photodetectors, solar cells and sensors, the performances of these devices suffer from unintentional oxygen impurities.[16, 22-27] Extensive studies on unintentional oxygen impurities in group III-nitrides showed a strong influence of surface oxide on the sensing parameters of nitride devices, apart from the variations in structural, electrical and optical properties.[22,23,28-32] Sensor device fabrication method itself has resulted in the metal-semiconductor junction with the presence of intermediate oxide layer. The gas sensitivity of such a device shows an overall enhancement by a factor of 50 as compared to that of the devices made without an oxide layer.[28] The reported observations of sensing response by III-nitride nanowires even without the assistance of any metal particles suggest the role of unintentional oxygen impurities in the chemical interactions at the semiconductor surface.[28,33,34] However, the underlying mechanism is yet to be understood.

During the sensing process, the chemisorbed molecules on a semiconductor surface induce a bending of energy bands near the semiconductor surface.[35] This surface band bending (SBB) can be precisely studied by the scanning Kelvin probe microscopy (SKPM). Apart from the type of semiconductor, the SBB (upward for *n*-type and downward for *p*-type) is strongly influenced by intrinsic and extrinsic defects, particle size, surface states, adsorbed gas or chemical molecules, incident photons, crystallographic orientations, surface reconstruction and temperature during the measurements.[35-39] The SKPM method can therefore be utilized for understanding the sensing mechanism involving chemisorption process.

The detection of potential greenhouse gas methane ($CH_4$) is extremely demanding in environmental safety.[40] In this study, we report the underlying mechanism of $CH_4$ sensing by GaN nanowires and the role of surface defects formed by the oxygen impurities in the nanowires.

The presence of oxygen impurity and the type of possible oxygen defects in nanowires are identified by the electron energy loss spectroscopic (EELS) studies. In this process, the EELS spectra corresponding to known oxygen defect complexes in GaN are simulated. These results are compared with the observed experimental data to identify the type of oxygen defect complexes present in the GaN nanowires. The gas sensing responses of the as-grown GaN nanowires are tested by fabricating ensemble nanowire sensor devices, without any metal catalyst decoration on the nanowires. In order to understand the charge transfer process during the gas sensing, *in situ* SKPM measurements are carried out on single GaN nanowires exposed under $CH_4$ gas.

## 2. Experimental details

*2.1 Nanowires synthesis and characterization*

Nonpolar GaN nanowires having different oxygen concentrations were synthesized using atmospheric pressure chemical vapor deposition technique in a vapor-liquid-solid (VLS) process. A detailed growth process, structural, compositional and optical study of the GaN nanowires with different oxygen impurity concentrations can be found elsewhere.[29] Four nanowire samples were grown under different oxygen impurity concentrations of around $10^5$ ppm (sample R1), $10^3$ ppm (sample R2), $10^2$ ppm (sample R3) and < 2 ppm (sample R4), for the present study.

Morphological features of the as-grown samples were examined by a field emission scanning electron microscope (FESEM, SUPRA 55 Zeiss). The EELS experiments were carried out to identify the presence of N and O in nanowires using an in-column second-order corrected Omega energy filter type spectrometer with an energy resolution of 0.7 eV in a high resolution transmission electron microscope (HRTEM, LIBRA 200FE Zeiss). The EELS calculations were performed by the density functional theory (DFT) based Cambridge Sequential Total Energy Package (CASTEP) code in the Materials Studio (accelrys) package.

*2.2 Nanowire device fabrication*

The ensemble GaN nanowire sensor devices were fabricated by depositing interdigitated electrodes of Ti/Al/Ti/Au metal films on the mats of as-grown nanowires, at ~$10^{-6}$ mbar using the thermal evaporation technique. The Ohmic nature of the contacts was tested after annealing the devices at 400 °C for 5 min in an inert $N_2$ atmosphere (supplementary information, Fig. S1). The pristine nanowires were mechanically dry transferred from the nanowires grown substrate to Au coated (~100 nm) conducting substrate for the SKPM measurements at room temperature as well as *in situ* measurements under gas exposure at elevated temperature.

*2.3 Sensing measurements of nanowire devices*

Agilent source measurement unit (B2902A) was used to study the current–voltage (I-V) characteristics of the fabricated ensemble nanowire devices. Gas sensing by four devices made of nanowire samples R1, R2, R3, and R4 were carried out using a custom built the in-house gas exposure facility under dynamic condition.[41] In the dynamic sensing, the gas was flown at a fixed flow rate over the sensor device while maintaining a fixed pressure in the sensing chamber under constant pumping using a dry pump. In the present case, responses of the devices were obtained at different concentrations and temperatures in the $N_2$ (99.999%) background at $10^{-2}$ mbar base pressure of the sensor chamber.

*2.4 SKPM measurements*

The SKPM studies were carried out for measuring the surface potential (SP) or the contact potential difference (CPD) on the nanowires at room temperature using an Agilent 5500 with a three-lock-in amplifier setup. For the measurements, a VAC bias at frequencies in the range 10-15 kHz plus a DC bias, VDC, were applied between an electrically conductive tip (SCM-PIT; Pt-Ir coated Si tip). The samples in a controlled environmental chamber with humidity level below 15% were maintained by continuously flowing pure $N_2$.[42]

The *in situ* localized gas exposure studies at elevated temperature of 100 °C on single nanowires under vacuum were studied using the Nanonics Multiprobe (MultiView 4000) imaging system. Simultaneous topography and SP maps were acquired by using a two-lock-in amplifier setup. The first lock-in amplifier recorded the fundamental resonance frequency (f ~ 68 KHz) of Si probe and the second lock-in amplifier used the second order mechanical resonance frequency (~ 420 KHz) of the Si probe. For localized gas delivery of few pico-liters ($10^{-12}$ liters) of gas on a single nanowire, Nanonics (nonoptical normal-force feedback) tuning fork based ($f_z$ ~34.4 kHz and Q-factor ~ 1080) FountainPen nano-pipette gas delivery probe with aperture size ~ 200 nm was used. Thus, a dynamic condition was also simulated here as only few pico-liters of gas were delivered in the large chamber volume of 1.5 liters.

**3. Results and discussions**

*3.1 Morphological and structural characteristics*

The morphologies of typical as-grown nanowires corresponding to samples R1 (Fig. 1(a)), R2 (Fig. 1(b)), R3 (Fig. 1(c)) and R4 (Fig. 1(d)) were studied. Nanowires R1 (Fig. 1(a)) and R2 (Fig. 1(b)), grown under oxygen rich atmospheres of $10^5$ and $10^3$ ppm, respectively, were found to be quite rough with non-uniform surface morphologies along the wires. These nanowires have a large diameter distribution of 40 – 150 nm. The diameter distribution of the nanowires of sample R3 (Fig. 1(c)), grown under the low oxygen concentration of $10^2$ ppm, is reduced to 40 – 100 nm. Nanowires R4 (Fig. 1(d)), grown under oxygen deficient atmospheres of < 2 ppm, showed uniform shape with homogeneous surface morphology and a narrow diameter distribution of ~ 60(±15) nm. The particle at the tip suggests that the growth of nanowires followed the VLS process. Detailed structural information of the nanowire samples R1 and R4 are discussed in the supplementary information (Fig. S4).

*3.2 Oxygen defects analysis by EELS spectra*

Previous investigations on oxygen defects in GaN revealed that the most favorable defect states are 2($O_N$) and $V_{Ga}$–3($O_N$).[31,32,43,44] Existence of the oxygen impurity and the aforementioned lattice oxygen defects in nanowire samples were investigated by comparing the simulated and experimental EELS spectra of GaN nanowires. It is worth mentioning that the EELS is highly sensitive to the presence of light elements as well as oxidation states. The experimental EELS spectra of all the nanowire samples were analyzed by collecting the *K*-edge emissions of N and O atoms. All the experimental core-loss spectra were background subtracted based on power law energy dependence ($AE^{-r}$). The relative concentration of N to O atoms ($n_N/n_O$) in the nanowires was estimated from the experimental EELS spectra. The $n_N/n_O$ is found to be increased from oxygen rich (0.3 for R1) to oxygen deficient nanowires (1.2 for R4).[29] The N *K*-edge and O *K*-edge EELS spectra corresponding to 2($O_N$) and $V_{Ga}$–3($O_N$) defects in GaN were then simulated and subsequently compared with the experimental spectra of GaN nanowires grown under different oxygen impurity concentrations.

The DFT+U calculations of the N *K*-edge and O *K*-edge EELS spectra were performed using the CASTEP code in the generalized gradient approximation (GGA) and the projector augmented wave (PAW) method. Experimental lattice parameters of wurtzite GaN, *a* = 3.189, and *c/a* = 1.626 were used for the calculations. The Monkhorst – Pack *k*-point grid of 8 × 8 × 4 and a plane wave energy cutoff value of 500 eV were chosen. A 2 × 2 × 2 supercell with 32 atoms was considered. The $O_N$ was created by replacing N with O atom, and $V_{Ga}$ was created by removing one Ga atom in the GaN supercell. A core-hole was introduced in the 1*s* orbital of N and O atoms to calculate respective core-loss *K*-edge spectra. The N *K*-edge spectra for a simulated GaN structure with 2($O_N$) and $V_{Ga}$–3($O_N$) defects were calculated by introducing a core-hole in N atom situated in the vicinity of the defect. In a DFT based EELS calculation, the calculated ionization edge and energy loss near edge spectroscopy (ELNES) features of an atom present in the vicinity of a defect dependends on the core-hole interaction.[45] Thus, it is sufficient enough to choose the size of the supercell larger than the unit cell of the crystal to avoid the core-hole interaction. Usually the 32- or 64-atoms supercell is sufficient for avoiding the core-hole interactions in the calculation of the EELS spectra, depending on the compound. We may also like to mention that, in order to calculate the accurate energy levels induced by the defects and to understand the charge transfer process, it is necessary to model the defects in GaN with larger size of supercell. Since we used 32 atom supercell especially for EELS calculation, any attempt to calculate the charge transfer process in individual defects to compare with the experimental SKPM data would be erroneous.

A pure GaN structure projected along Z-axis ∥ (0001) is shown in Fig. 2(a). The simulated N *K*-edge spectrum of pure GaN system is compared with the experimental N *K*-edge spectrum of sample R4 having lowest oxygen concentration (Fig. 2(a)). Both the experimental and simulated spectra matched well except for a small energy shift among the extended energy loss featuring around 425 eV. The ionization edge of N *K*-edge spectrum is found to be around 399 eV. The calculated N *K*-edge spectrum for the GaN structure with 2($O_N$) defect was found to match with the experimental spectrum (Fig. 2(b)) of R3. Both the ionization edges and the near-edge fine structure features matched well. Similarly, the O K-edge spectrum of the simulated $2(O_N)$ defect state in GaN was compared with that of the experimental data of R3 (Fig. 2(b)). The experimental spectrum is quite broad compared to the calculated spectrum. However, the near-edge and the extended energy loss features resembled that of the experimental data. The ionization edge of the O *K*-edge spectra is found to be around 531 eV.

Similarly, the *K*-edge spectrum (middle of Fig. 2(c)) was calculated for the N atom in the vicinity of $V_{Ga}$–$3(O_N)$ defect configuration in GaN structure and was found to match closely with the experimental spectrum of the nanowires of R1 including the near-edge features. The calculated O *K*-edge spectrum (bottom of Fig. 2(c)) of the $V_{Ga}$–$3(O_N)$ defect also matched well with that of the sample R1. It is noteworthy to mention that the clear near-edge fine structures of N and O *K*-edges (Fig. 2(c)) are observed for both the experimental and the simulated spectra. Thus the EELS study and simulations simulation helped in understanding the existence of the $2(O_N)$ and $V_{Ga}$–$3(O_N)$ oxygen defects in GaN nanowires and also provided insights into their dependence on the concentration of oxygen impurity during the nanowire growth. Earlier study based on first-principles total energy calculations for the oxygen defects in GaN (surface oxygen coverage of about $10^{15}$ cm$^{-2}$) showed that $2(O_N)$ and $V_{Ga}$–$3(O_N)$ defect configurations had low surface energies [0.5 – 1.7 eV/ (1 × 1) over a range of Ga chemical potential] in comparison to other defect states.[43] Among the two lattice defect states, $2(O_N)$ is the stable configuration with low surface energy as compared to that of $V_{Ga}$–$3(O_N)$.

*3.3 Gas sensing characteristics*

Gas sensing properties for the four sensor devices made of the nanowire samples R1, R2, R3, and R4 were studied towards $CH_4$ (99.999%), a reducing gas. $CH_4$ is chosen ahead of the reducing gasses like inflammable $H_2$ and corrosive $NH_3$ to have an ease in the handling, particularly, due to the limitation of the SKPM chamber for the *in situ* measurement using these gases. The sensor response transients at different concentrations (50–500 ppm) of $CH_4$ and at different temperatures (50–150 °C) of the devices were recorded. The devices of sample R1, R2 and R3 showed a significant response at 125 °C even at low concentration of 50 ppm gas exposure (Figs. 3(a), 3(b) and 3(c)). In contrast, the device of sample R4 (Fig. 3(d)) did not respond even for 500 ppm. The sensor responses ($R_N$ - $R_G$ / $R_N$) ×100 % where $R_N$ and $R_G$ represent the resistances of the devices in $N_2$ background and sensing gas, respectively, are calculated. The maximum response at 500 ppm of $CH_4$ for the sensor devices of samples R1, R2 and R3 are 1.27 %, 0.87%, and 0.54 %, respectively. The gas responses of the devices were reproducible even after three months from the fabrication date. The responses of these devices are of similar order to that for the reported values in case of the $CH_4$ sensing by GaN and $SnO_2$ nanowires.[12,16,34]

The three devices showed a linear response over a concentration range of 50 – 500 ppm of $CH_4$ (Fig. S2). There is a gradual reduction in the sensing response as the oxygen concentration in the GaN NWs decreases from R1 to R3 (Fig. 3). Temperature dependent $CH_4$ sensing measurements showed increased amount of response with increasing temperature (Fig. S3). The lowest temperature at which the devices responded was 50 – 60 °C. The influence of temperature on the sensing response confirmed the chemisorption instead of physical adsorption of the $CH_4$ on nanowire surface.[11] The increase in the temperature provides an activation energy for chemisorption induced sensor responses leading to an effective and efficient use of surface binding sites. It might be noted that the chemisorption of $CH_4$ on the pure GaN surface did not occur even at high temperature without the assistance of a metal catalyst.[7,17] However, the observed responses by the nanowire samples suggested the presence of active sites which might have driven the chemisorption of $CH_4$. The active sites in GaN nanowires could be lattice oxygen defects, which might have formed during the nanowire growth. For understanding the type of oxygen defects responsible for sensing by nanowires, the results of EELS study were utilized. The experimental and simulated EELS studies revealed the existence of oxygen defects $2(O_N)$ and $V_{Ga}–3(O_N)$ in the nanowires. However, the presence of crystalline phase consisting of Ga and oxygen was ruled out from the Raman studies of ensemble and single GaN nanowire grown under oxygen rich condition (sample R1), as Raman modes corresponding to the oxide phase were not observed (Fig. S5).[46] Thus, oxygen may only be present as point defects or its complexes. During the sensing experiments, the $CH_4$ adsorption occurred strongly on the nanowire surfaces due to $2(O_N)$ and $V_{Ga}–3(O_N)$ defect states and the net charge transfer process took place between these defects and $CH_4$. The recovery process could therefore be explained by further adsorption of the residual oxygen, present in the exposure chamber, on the active sites involved in the sensing process.

*3.4 Surface potential measurements of as-grown nanowires*

The underlying mechanism of $CH_4$ sensing by GaN nanowires, with various concentrations of lattice oxygen defects, was probed by studying the SP or the CPD variations on several nanowire surfaces using the SKPM technique. In SKPM mode, usually the CPD is compensated by applying external bias voltages that facilitate a simultaneous acquisition of topography and the surface potential SP.[42] The CPD can be written as

$$V_{CPD} = (\phi_M - \phi_S) / q \qquad (1)$$

where $\phi_M$ and $\phi_S$ are work function of the metal tip and the sample, respectively, and $q$ is the elementary charge. Any change in the work function of the sample is reflected directly in the CPD values. As discussed earlier in the introduction, the SP measurements are influenced by several factors. The SP variations of the as-grown GaN nanowires and the influential factors such as the diameter and impurity concentration in nanowires were studied at room temperature prior to CH$_4$ exposure of the nanowires.

Simultaneous topography and CPD (SP) maps were acquired for the four nanowire samples R4 (Fig. 4(a)), R3 (Fig. 4(b)), R2 (Fig. 4(c)) and R1 (Fig. 4(d)). A typical topography map of nanowires R4 (Fig. 4(a), left side), grown under reduced oxygen atmosphere shows a quite smooth and homogeneous surface morphology along the nanowire having an average diameter ~ 60 nm. The corresponding SP map (Fig. 4(a), right side) depicts the uniform SP contrast with an average value of 86 (±2) mV, along the nanowire. As the concentration of the oxygen was increased during the growth (for samples R3, R2 and R1), the topography maps of nanowires R3 (Fig. 4(b), left side), R2 (Fig. 4(c), left side) and R1 (Fig. 4(a), left side) showed increased amount of surface roughness along the wires. The RMS roughness on these surfaces increased from ~ 8 to 26 nm for samples R4 to R1. The FESEM studies also qualitatively supported this observation (Fig. 1). The corresponding SP maps of nanowires R3 (Fig. 4(b), right side), R2 (Fig. 4(c), right side) and R1 (Fig. 4(d), right side) revealed the inhomogeneous SP contrast resembling topography variations along the surface of the nanowires. The SP variation could be due to different facets present in the rough morphologies.

A statistical analysis of SP values of nanowires R4 with different diameters was also carried out. The results showed (Fig. 4(e)) an increase in the SP values with increasing nanowire diameter. This observation is in accordance to SBB due to the defects as explained subsequently. GaN behaves as an *n*-type semiconductor due to the intrinsic N vacancies (V$_N$) and unintentional substituent O at N site (O$_N$).[47] The excess negative charges at the surface of an *n*-type semiconductor (GaN) cause the accumulation of positive charges near the surface leading to the formation of a space charge or the depletion region. In this case, the surface bands bend upward resulting to a positive SP value. With an increase in the nanowire diameter, the surface charge density reduces for a fixed amount of doping concentration in R4. This reduction in the surface charge density also diminishes the accumulation of positive charges near the surface. Consequently, the resultant SP value is found to increase with increasing diameter (Fig. 4(e)). Statistics of SP values (Fig. 4(f)) of similar diameter (~ 45 nm) nanowires with different oxygen concentration showed an increase in the SP value from 66 (±4) to 160 (±4) mV with increasing oxygen impurity in nanowires (from samples R4 to R1). The increase in the SP values (more positive) with oxygen impurity is evident from the fact that the increased majority carrier concentration by oxygen donors reduces the upward band bending. So the SP values became more positive. Apart from the dopant concentration, the net SBB observed in oxygen rich nanowires could be a result of the surface states and surface defects.

*3.5 in situ SKPM measurements under gas exposure*

The local charge transfer mechanism in the $CH_4$ sensing was probed by acquiring SP or CPD maps of the nanowires using SKPM with the *in situ* gas exposure multiprobe setup. A schematic view of the setup along with an optical image is shown (Fig. 5). The setup consisted of a standard beam-bounce feedback cantilever based Au coated Si probe (dashed red arrow in Fig. 5) for the SKPM measurements which were carried out in an intermittent mode of atomic force microscopy (AFM) with a tip-sample distance of ~20 nm. By positioning the Kelvin probe and the gas delivery nano-pipette probe (solid arrow line in Fig. 5) facing in opposite and close to each other (Fig. 5), under their respective feedback, simultaneous topography and CPD maps were acquired with sample stage scanning. This arrangement allows us to expose the nanowires to target gas locally, using a nano-pipette probe and to measure the CPD on nanowires using the SKPM probe.

Before proceeding to the SKPM measurements on the nanowire samples, Au coated Si tip was calibrated on pure Au film under similar conditions of the SKPM measurements under an absolute concentration of 1000 ppm $CH_4$ (99.999%) in $N_2$ (99.999 %) base gas delivered through the nano-pipette probe. The results (Fig. S6) showed that the tip work function did not change with the experimental conditions in the SKPM chamber. Prior to the *in situ* SKPM measurements at elevated temperature, the nanowire samples on Au substrate were kept in the SKPM chamber at ~$10^{-6}$ mbar pressure for 12 h in a dark condition and at room temperature. The samples were then heated to 125 °C for 2 h at ~$10^{-6}$ mbar pressure in dark and brought to the required temperature for the SKPM measurements and were allowed to stabilize for 2h. The procedure helped the samples to dehumidify and stabilize the surface potentials. Among the four samples, the SKPM measurements were carried out on two nanowire samples grown under two extreme cases of oxygen deficient (sample R4) and oxygen rich (sample R1) conditions by maintaining similar experimental conditions of 100 °C at ~$10^{-2}$ mbar base pressure, at which gas sensing experiments were performed. Simultaneous topography and CPD (or SP) maps were acquired before the $CH_4$ exposure as well as during the $CH_4$ exposure.

A typical topography map of nanowires R4 (Fig. 6(a)), and the corresponding CPD maps before (Fig. 6(b)) and during (Fig. 6(c)) the $CH_4$ exposure are displayed. A statistical analysis of CPD values as a function of the nanowire diameter, both before and during the $CH_4$ exposure showed (Fig. 6(g)) that the CPD value increased with the increasing nanowire diameter. The CPD value, before the $CH_4$ exposure is found to increase from ~70 to ~130 mV when the diameter of the nanowire is increased from ~ 40 to ~70 nm. Variation in the CPD values in comparison to the values obtained at room temperature and atmospheric pressure (Fig. 4(e)) are expected due to the difference in work functions of the conducting tips, temperature and base pressure of the SKPM chamber. However, the change in the CPD values due to $CH_4$ exposure [$\Delta V_{CPD} = V_{CPD}$ (before) − $V_{CPD}$ (during) $CH_4$], at a particular temperature and a base pressure with a fixed conducting tip was only considered. In the present study, the CPD values are reduced to a range of ~ 65 - 117 mV, during the $CH_4$ exposure. It is noteworthy to mention that the $\Delta V_{CPD}$ values are very low (in the range of 5 - 17 mV) during the $CH_4$ exposure for any size of the nanowire diameter for the sample R4.

Similarly, a typical topography map of nanowires for R1 (Fig. 6(d)), and the corresponding CPD maps before (Fig. 6(e)) and during (Fig. 6(f)) the $CH_4$ exposure were recorded at 100 °C and $10^{-2}$ mbar. Statistical analysis of CPD values before the $CH_4$ exposure (Fig. 6(h)), as a function of the nanowire diameter, showed the similar trend of increase (~ 176 to ~790 mV) with increasing nanowire diameter (~35 to ~130 nm). Whereas the CPD values lost correlation with nanowire diameter during the $CH_4$ exposure and fluctuated in the range of ~ 210 - 430 mV with a mean saturation value of about 330 (± 59) mV for the entire range of nanowire diameter. Considering the saturation mean value of CPD (330 (± 59) mV) during the $CH_4$ exposure, we obtained the $\Delta V_{CPD}$ values before and during the gas exposure. The value of $\Delta V_{CPD}$ is found to be positive (+ 460 mV) for higher diameter nanowires while it is negative ($\Delta V_{CPD}$ = −154 mV) for nanowires with the smaller diameter. The changes in the CPD values, during the $CH_4$ exposure, are in the range of ~ 150 - 460 mV. Thus, the $\Delta V_{CPD}$ of nanowires R1, grown under oxygen rich atmosphere, is one order higher than that of R4 (5 - 17 mV), grown under reduced oxygen atmosphere.

The SBB due to the CH$_4$ exposure on GaN nanowires of R1 was calculated by considering the highest possible amount of oxygen dopants of ~ 10$^{21}$ cm$^{-3}$ in GaN lattice.[31] The SBB in dark can be written as

$$\Phi_B = \phi_M - qV_{CPD} + \phi_{off} - \chi + (E_c - E_f) \quad (2)$$

where $\phi_{off}$ is the offset in work function of the metal tip found by calibration (~ 28 meV) on Au film ($\phi_M$ = 5.1 eV) and $\chi$ is electron affinity of the GaN (3.2 eV).[39] $E_c$ and $E_f$ are positions of conduction band minimum and Fermi level, respectively. The estimated SBB in dark at 100 °C under 10$^{-2}$ mbar base pressure, by measuring V$_{CPD}$ before the CH$_4$ exposure, is found to be 1.57 - 0.95 eV over a range of nanowire diameters of 35 - 130 nm (Fig. 7(a)). The space charge region formed due to the SBB is characterized by the depletion width, $W$. The standard expression assuming the semi-infinite configuration instead of cylindrical geometry is chosen to calculate the depletion width considering the high surface roughness (~26 nm) which is a significant fraction of the nanowire diameter with non-uniform shape and surface morphology.[48] Therefore,

$$W = (n_s / N_D) = (2 \Phi_B \varepsilon \varepsilon_0 / q^2 N_D)^{1/2} \quad (3)$$

where $\varepsilon$ is the static dielectric constant of GaN (9.8),[39] $\varepsilon_0$ is the permittivity of free space, $N_D$ is dopant concentration, and $n_s$ is surface charge density.

The calculated depletion width ($W$) (Fig. 7(b)) and the corresponding surface charge density ($n_s$) (Fig. 7(c)), before the gas exposure are 12.9 to 10.1 nm and 1.29 × 10$^{15}$ to 1.01 × 10$^{15}$ cm$^{-2}$, respectively. During the CH$_4$ exposure, the estimated SBB ($\Phi_B$) is found to fluctuate and saturate on a mean value of 1.41 (± 0.06) eV over the range of nanowire diameters of 35 – 130 nm (Fig. 7(a)). Thus, $W$ and $n_s$ saturated at about the mean values of 12.3 (± 0.3) nm and 1.23 × 10$^{15}$ cm$^{-2}$, respectively following the eq. 3 (Figs. 7(b) and 7(c)). The saturated mean values showed an increase in $\Phi_B$, $W$ and $n_s$ for nanowires with higher diameter when they were compared with the estimated values before the $CH_4$ exposure. The $\Phi_B$ and $n_s$ due to the $CH_4$ exposure are increased by an amount of 0.46 eV and $2.2 \times 10^{14}$ cm$^{-2}$. As the nanowire diameter decreased, the change in SBB ($\Delta\Phi_B$), $W$ ($\Delta W$), and $n_s$ ($\Delta n_s$) values also got diminished. Below a critical diameter of ~ 60 nm the trend of decrease in the $\Phi_B$, $W$, and $n_s$ is observed with respect to the mean value, as compared with the estimated values before the $CH_4$ exposure. The $\Phi_B$ and $n_s$ due to the $CH_4$ exposure are decreased by an amount of 0.16 eV and $6.0 \times 10^{13}$ cm$^{-2}$, for nanowires with the smaller diameter. A large number of nanowires with small diameter (<60 nm) are present in the oxygen rich sample (R1; Fig. S7).

*3.6 Charge transfer and surface band bending mechanism*

The estimated changes in $\Phi_B$, $W$, and $n_s$ values were analyzed to understand the gas sensing mechanism in our chemiresistive measurements (Fig. 3). During the gas exposure, a charge transfer occurs between adsorbed $CH_4$ and the nanowire surface which have active oxygen defect complexes of $2(O_N)$ and $V_{Ga}$–$3O_N$ (Fig. 7(d)). As discussed earlier, $2(O_N)$ defects are more stable than the high energetic $V_{Ga}$–$3O_N$ defects.[43] In such case, the $2(O_N)$ defects are likely to be prevalent in the nanowires with higher diameter (with lower surface-to-volume ratio) than that of the smaller ones. Due to the dominant presence of stable $2(O_N)$ defects on the surface of the nanowire with higher diameter, the excess negative charges were therefore transferred by $CH_4$ accumulated on the surfaces. Thus, the surface charge density, $n_s$ got increased (Fig. 7(c)) and it helped the surface bands to bend further upward from 0.95 to 1.41 ($\pm$ 0.06) eV (Fig. 7(a)). As a result, the depletion width ($W$) was found to increase in nanowires with higher diameter (Fig. 7(b)).

A large surface energy is expected for the smaller diameter nanowires because of their high surface-to-volume ratio in addition to the surface roughness (~26 nm for sample R1), which is a significant fraction for the case of small diameters. Calculated surface-to-volume ratio for the sample with rough (R1 with RMS roughness 26 nm) and smooth (R4 with RMS roughness 8 nm) surfaces for an area of 76 × 76 nm² are found to be ~165 and ~15, respectively. Thus, the high energetic $V_{Ga}$–$3O_N$ defect complexes are ubiquitous on the surface of oxygen rich nanowires of smaller diameter with high surface roughness as observed for the oxygen rich sample R1. In this context, it should be noted that the surface roughness only contributes to the enhancement of sensitivity of gas sensing by increasing the effective area and does not contribute for the charge transfer process. Because of the dominant presence of less stable $V_{Ga}$–$3O_N$ defects, the lattice oxygen available in the defect complexes with sufficiently high surface energy participated in the electron transfer process with the $CH_4$. Consequently, oxygen from the $V_{Ga}$–$3O_N$ complexes released out after receiving an electron from the $CH_4$ (Fig. 7(d)). So the net surface charge density, $n_s$ was reduced resulting in the reduction of both SBB and depletion width values (Fig. 7).

Although the calculated change in the local depletion width of smaller diameter nanowires is low, the global change of depletion width is expected to be high due to the effective surfaces exposed to the $CH_4$ are large for smaller diameter nanowires. As a result of the reduction in the depletion width, the conduction path of the majority carriers in the nanowire was enhanced leading to the improved carrier transportation. This process contributed to the resistive gas sensing measurement by lowering the total resistance during the $CH_4$ exposure. When the smaller diameter nanowires with $V_{Ga}$–$3O_N$ surface defects are large in number, the corresponding resistance due to $CH_4$ exposure is found to decrease during the gas sensing measurements. The localized charge transfer process involving $V_{Ga}$–$3O_N$ defects, as observed in single nanowires, may thus be responsible in controlling the global gas sensing for the oxygen rich ensemble of GaN nanowires. Notably a decrease in the resistance is observed (Fig. 3) for the oxygen rich sample during the $CH_4$ exposure. When the $CH_4$ flow was switched off, the oxygen atoms present in the sensing atmosphere get adsorbed on the vacant $O_N$ site resulting in the increase of the depletion width. As a result, the resistance of the devices recovers to the base level.

## 4. Summary

The resistive gas sensor devices made of GaN nanowires with different oxygen impurity concentrations showed the response of reduction in resistance, towards the exposure with $CH_4$. The experimental and the simulation of EELS studies on oxygen impurities in GaN nanowires confirmed the possible presence of $2(O_N)$ and $V_{Ga}$–$3O_N$ defect complexes. The SKPM measurements showed a decrease in the SBB value with an increase in the diameter. A reduction in the SBB value is also observed as the oxygen concentration in nanowires increases. The observed variations in SBB, depletion width, and surface charge density during the SKPM measurements on nanowires exposed to $CH_4$ confirmed the occurrence of gas adsorption and charge transfer processes in these nanowires. A localized charge transfer process, involving $V_{Ga}$–$3O_N$ defect complex in nanowires is attributed in controlling the global gas sensing behavior of the oxygen rich GaN nanowire ensemble. Thus, by tuning the surface lattice oxygen defects on nanowires, modulation of the sensing behavior of GaN nanowires is demonstrated. The present results provide an insight for controlling the defects in III-nitride based nanostructures for advanced sensor device applications.

**Acknowledgements :**

One of us (AP) thanks the Department of Atomic Energy, India for financial aid. The authors acknowledge the expertise of R. Pandian, SND for FESEM studies. We also thank S. Chandra, and P. Jagadeesan of MPD, IGCAR for useful discussions in the simulation studies. The UGC-DAE CSR, Kalpakkam node is acknowledged for the computation and electron microscopy facility.

**Supporting Information Available:**

Electrical characterizations, Structural information by TEM, SKPM tip calibration, temperature dependent sensor response, Raman spectroscopic studies and nanowire diameter distribution. This material is available free of charge via the Internet at http://pubs.acs.org.

**References :**

1. Kibria, M. G.; Zhao, S.; Chowdhury, F. A.; Wang, Q.; Nguyen, H. P.; Trudeau, M. L.; Guo, H.; Mi, Z. Tuning the surface Fermi level on *p*-type gallium nitride nanowires for efficient overall water splitting. *Nat. Commun.* **2014**, *5*, 3825.

2. AlOtaibi, B.; Nguyen, H. P.; Zhao, S.; Kibria, M. G.; Fan, S.; Mi, Z. Highly stable photoelectrochemical water splitting and hydrogen generation using a double-band InGaN/GaN core/shell nanowire photoanode. *Nano Lett.* **2013**, *13*, 4356-4361.

3. Wang, D.; Pierre, A.; Kibria, M. G.; Cui, K.; Han, X.; Bevan, K. H.; Guo, H.; Paradis, S.; Hakima, A. R.; Mi, Z. Wafer-level photocatalytic water splitting on GaN nanowire arrays grown by molecular beam epitaxy. *Nano Lett.* **2011**, *11*, 2353-2357.

4. Shen, X.; Small, Y. A.; Wang, J.; Allen, P. B.; Fernandez-Serra, M. V.; Hybertsen, M. S.; Muckerman, J. T., Photocatalytic water oxidation at the GaN (10-10) − water interface. *J. Phys. Chem. C* **2010**, *114*, 13695-13704.

5. Chen, C. P.; Ganguly, A.; Lu, C. Y.; Chen, T. Y.; Kuo, C. C.; Chen, R. S.; Tu, W. H.; Fischer, W. B.; Chen, K. H.; Chen, L. C. Ultrasensitive in situ label-free DNA detection using a GaN nanowire-based extended-gate field-effect-transistor sensor. *Anal. Chem.* **2011**, *83*, 1938-1943.

6. Cui, Y.; Wei, Q.; Park, H.; Lieber, C. M. Nanowire nanosensors for highly sensitive and selective detection of biological and chemical species. *Science* **2001**, *293*, 1289-1292.

7. Dobrokhotov, V.; McIlroy, D. N.; Norton, M. G.; Abuzir, A.; Yeh, W. J.; Stevenson, I.; Pouy, R.; Bochenek, J.; Cartwright, M.; Wang, L.; *et. al.* Principles and mechanisms of gas sensing by GaN nanowires functionalized with gold nanoparticles. *J. Appl. Phys.* **2006**, *99*, 104302.

8. Göpel, W.; Schierbaum, K. D. $SnO_2$ sensors: current status and future prospects. *Sensors and Actuators B* **1995**, *26*, 1-12.

9. Wang, B.; Zhu, L. F.; Yang, Y. H.; Xu, N. S.; Yang, G. W., Fabrication of a $SnO_2$ nanowire gas sensor and sensor performance for hydrogen. *J. Phys. Chem. C* **2008**, *112*, 6643-6647.

10. Antony, R. P.; Mathews, T.; Dash, S.; Tyagi, A., Kinetics and physicochemical process of photoinduced hydrophobic ↔ superhydrophilic switching of pristine and N-doped $TiO_2$ nanotube arrays. *J. Phys. Chem. C* **2013**, *117*, 6851-6860.

11. Batzill, M.; Diebold, U. The surface and materials science of tin oxide. *Prog. Surf. Sci.* **2005**, *79*, 47-154.

12. Bonu, V.; Das, A.; Prasad, A. K.; Krishna, N. G.; Dhara, S.; Tyagi, A. K. Influence of in-plane and bridging oxygen vacancies of $SnO_2$ nanostructures on $CH_4$ sensing at low operating temperatures. *Appl. Phys. Lett.* **2014**, *105*, 243102.

13. Kohl, D. The role of noble metals in the chemistry of solid-state gas sensors. *Sensors and Actuators B* **1990**, *1*, 158-165.

14. Hazra, A.; Dutta, K.; Bhowmik, B.; Chattopadhyay, P.; Bhattacharyya, P. Room temperature alcohol sensing by oxygen vacancy controlled $TiO_2$ nanotube array. *Appl. Phys. Lett.* **2014**, *105*, 081604.

15. Sahoo, P.; Dhara, S.; Dash, S.; Amirthapandian, S.; Prasad, A. K.; Tyagi, A. K. Room temperature $H_2$ sensing using functionalized GaN nanotubes with ultra low activation energy. *Int. J. Hydrogen Energy* **2013**, *38*, 3513-3520.

16. Lim, W.; Wright, J. S.; Gila, B. P.; Johnson, J. L.; Ural, A.; Anderson, T.; Ren, F.; Pearton, S. J. Room temperature hydrogen detection using Pd-coated GaN nanowires. *Appl. Phys. Lett.* **2008**, *93*, 072109.

17. Pearton, S.; Ren, F. Gallium nitride-based gas, chemical and biomedical sensors. *IEEE T Instrum. Meas.* **2012**, *15*, 16-21.

18. Baik, K. H.; Kim, H.; Lee, S.-N.; Lim, E.; Pearton, S.; Ren, F.; Jang, S. Hydrogen sensing characteristics of semipolar (11-22) GaN Schottky diodes. *Appl. Phys. Lett.* **2014**, *104*, 072103.

19. Sahoo, P.; Suresh, S.; Dhara, S.; Saini, G.; Rangarajan, S.; Tyagi, A. K. Direct label free ultrasensitive impedimetric DNA biosensor using dendrimer functionalized GaN nanowires. *Biosens. Bioelectron.* **2013**, *44*, 164-170.

20. Lo, C.; Chang, C.; Chu, B.; Pearton, S.; Dabiran, A.; Chow, P.; Ren, F. Effect of humidity on hydrogen sensitivity of Pt-gated AlGaN/GaN high electron mobility transistor based sensors. *Appl. Phys. Lett.* **2010**, *96*, 232106.

21. Aluri, G. S.; Motayed, A.; Davydov, A. V.; Oleshko, V. P.; Bertness, K. A.; Sanford, N. A.; Rao, M. V. Highly selective GaN-nanowire/$TiO_2$-nanocluster hybrid sensors for detection of benzene and related environment pollutants. *Nanotechnology* **2011**, *22*, 295503.

22. Pearton, S. J.; Zolper, J. C.; Shul, R. J.; Ren, F. GaN: Processing, defects, and devices. *J. Appl. Phys.* **1999**, *86*, 1.

23. Toth, M.; Fleischer, K.; Phillips, M. Direct experimental evidence for the role of oxygen in the luminescent properties of GaN. *Phys. Rev. B* **1999**, *59*, 1575-1578.

24. Wu, J. When group-III nitrides go infrared: New properties and perspectives. J. *Appl. Phys.* **2009**, *106*, 011101.

25. Zhong, Z.; Qian, F.; Wang, D.; Lieber, C. M. Synthesis of *p*-Type Gallium nitride nanowires for electronic and photonic nanodevices. *Nano Lett.* **2003**, *3*, 343-346.

26. Gonzalez-Posada, F.; Songmuang, R.; Den Hertog, M.; Monroy, E. Room-temperature photodetection dynamics of single GaN nanowires. *Nano Lett.* **2012**, *12*, 172-176.

27. Choi, H.-J.; Johnson, J. C.; He, R.; Lee, S.-K.; Kim, F.; Pauzauskie, P.; Goldberger, J.; Saykally, R. J.; Yang, P., Self-Organized GaN quantum wire UV lasers. *J. Phys. Chem. B* **2003**, *107*, 8721-8725.

28. Weidemann, O.; Hermann, M.; Steinhoff, G.; Wingbrant, H.; Lloyd Spetz, A.; Stutzmann, M.; Eickhoff, M. Influence of surface oxides on hydrogen-sensitive Pd:GaN Schottky diodes. *Appl. Phys. Lett.* **2003**, *83*, 773-775.

29. Patsha, A.; Amirthapandian, S.; Pandian, R.; Dhara, S. Influence of oxygen in architecting large scale nonpolar GaN nanowires. *J. Mater. Chem. C* **2013**, *1*, 8086-8093.

30. Arslan, I.; Browning, N. D. Role of oxygen at screw dislocations in GaN. *Phys. Rev. Lett.* **2003**, *91*, 165501.

31. Slack, G. A.; Schowalter, L. J.; Morelli, D.; Freitas Jr, J. A. Some effects of oxygen impurities on AlN and GaN. *J. Cryst. Growth* **2002**, *246*, 287-298.

32. Reshchikov, M. A.; Morkoç, H. Luminescence properties of defects in GaN. *J. Appl. Phys.* **2005**, *97*, 061301-061395.

33. Yun, F.; Chevtchenko, S.; Moon, Y.-T.; Lnq jnb, H.; Fawcett, T. J.; Wolan, J. T. GaN resistive hydrogen gas sensors. *Appl. Phys. Lett.* **2005**, *87*, 073507.

34. Popa, V.; Tiginyanu, I.; Ursaki, V.; Volcius, O.; Morkoç, H. A GaN-based two-sensor array for methane detection in an ethanol environment. *Semicond. Sci. Technol.* **2006**, *21*, 1518-1521.

35. Zhang, Z.; Yates Jr, J. T. Band bending in semiconductors: chemical and physical consequences at surfaces and interfaces. *Chem. Rev.* **2012**, *112*, 5520-5551.

36. Barbet, S.; Aubry, R.; di Forte-Poisson, M. A.; Jacquet, J. C.; Deresmes, D.; Lḱm S-: Sgḋron, D. Surface potential of *n*- and *p*-type GaN measured by Kelvin force microscopy. *Appl. Phys. Lett.* **2008**, *93*, 212107.

37. Soudi, A.; Hsu, C.-H.; Gu, Y. Diameter-dependent surface photovoltage and surface state density in single semiconductor nanowires. *Nano Lett.* **2012**, *12*, 5111-5116.

38. Chevtchenko, S.; Ni, X.; Fan, Q.; Baski, A.; Morkoc, H. Surface band bending of *a*-plane GaN studied by scanning Kelvin probe microscopy. *Appl. Phys. Lett.* **2006**, *88*, 122104.

39. Foussekis, M.; McNamara, J. D.; Baski, A. A.; Reshchikov, M. A. Temperature-dependent Kelvin probe measurements of band bending in *p*-type GaN. *Appl. Phys. Lett.* **2012**, *101*, 082104.

40. Yvon-Durocher, G.; Allen, A. P.; Bastviken, D.; Conrad, R.; Gudasz, C.; St-Pierre, A.; Thanh-Duc, N.; Del Giorgio, P. A. Methane fluxes show consistent temperature dependence across microbial to ecosystem scales. *Nature* **2014**, *507*, 488-491.

41. Prasad, A. K.; Amirthapandian, S.; Dhara, S.; Dash, S.; Murali, N.; Tyagi, A. K. Novel single phase vanadium dioxide nanostructured films for methane sensing near room temperature. *Sensors and Actuators B* **2014**, *191*, 252-256.

42. Sahoo, P.; Oliveira, D. S.; Cotta, M. A.; Dhara, S.; Dash, S.; Tyagi, A. K.; Raj, B. Enhanced surface potential variation on nanoprotrusions of GaN microbelt as a probe for humidity sensing. *J. Phys. Chem. C* **2011**, *115*, 5863-5867.

43. Northrup, J. E. Oxygen-rich GaN (10-10) surfaces: First-principles total energy calculations. *Phys. Rev. B* **2006**, *73*, 115304.

44. Elsner, J.; Jones, R.; Haugk, M.; Gutierrez, R.; Frauenheim, T.; Heggie, M.; Oberg, S.; Briddon, P. Effect of oxygen on the growth of (10-10) GaN surfaces: The formation of nanopipes. *Appl. Phys. Lett.* **1998**, *73*, 3530-3532.

45. Egerton, R. Electron energy-loss spectroscopy in the electron microscope, *Springer Science+ Business Media* **2011**.

46. Lan, Z.-H.; Liang, C.-H.; Hsu, C.-W.; Wu, C.-T.; Lin, H.-M.; Dhara, S.; Chen, K.-H.; Chen, L.-C.; Chen, C.-C. Nanohomojunction (GaN) and nanoheterojunction (InN) nanorods on one-dimensional GaN nanowire substrates. *Adv. Funct. Mater.* **2004**, *14*, 233-237.

47. Zhu, T.; Oliver, R. A., Unintentional doping in GaN. *Phys. Chem. Chem. Phys.* **2012**, *14*, 9558-9573.

48. Calarco, R.; Marso, M.; Richter, T.; Aykanat, A. I.; Meijers, R.; A, V. D. H.; Stoica, T.; Luth, H. Size-dependent photoconductivity in MBE-grown GaN-nanowires. *Nano Lett.* **2005**, *5*, 981-984.

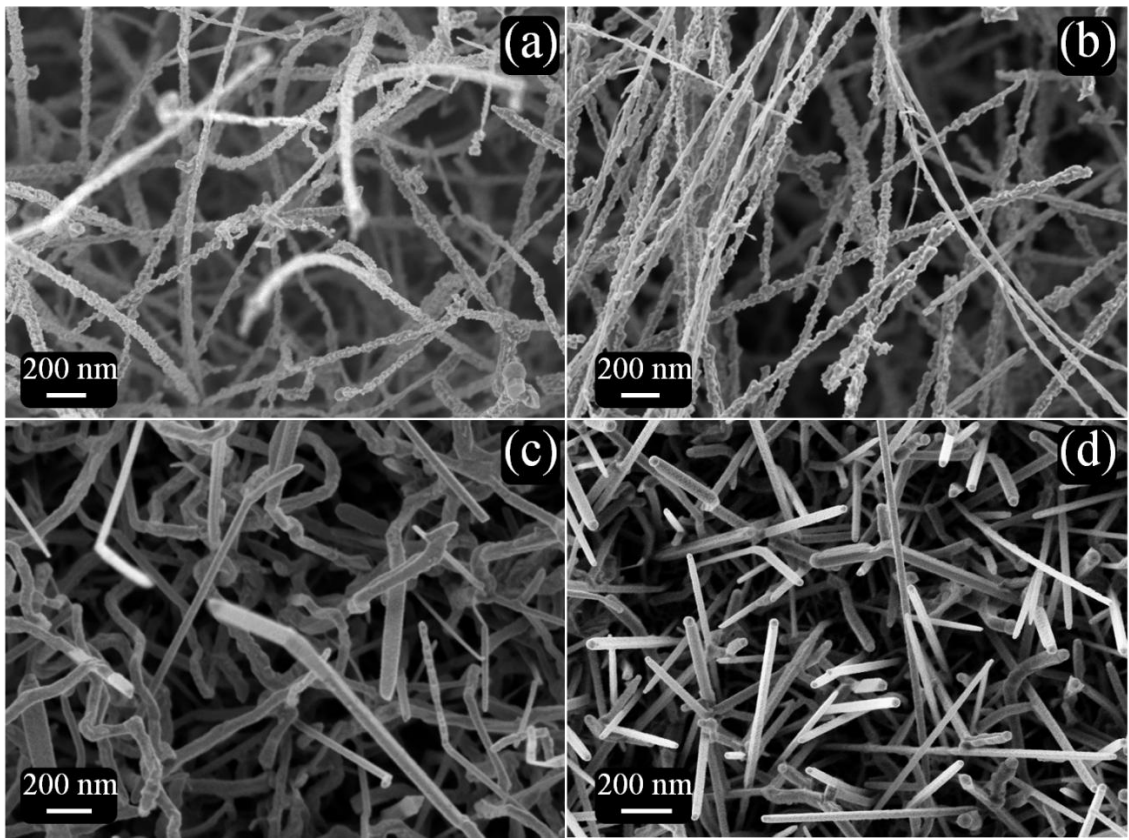

**Figure 1**. Typical FESEM micrographs of GaN nanowires grown under different oxygen concentrations of (a) ~$10^5$ ppm; sample R1, (b) ~$10^3$ ppm; sample R2. Both show quite rough and nonuniform surface morphology. (c) Nanowires grown under oxygen concentrations of ~$10^2$ ppm; sample R3 and (d) < 2 ppm; sample R4, having the uniform diameter and surface morphology along the nanowires. Nanoparticles at the tip of the nanowires are clearly observed.

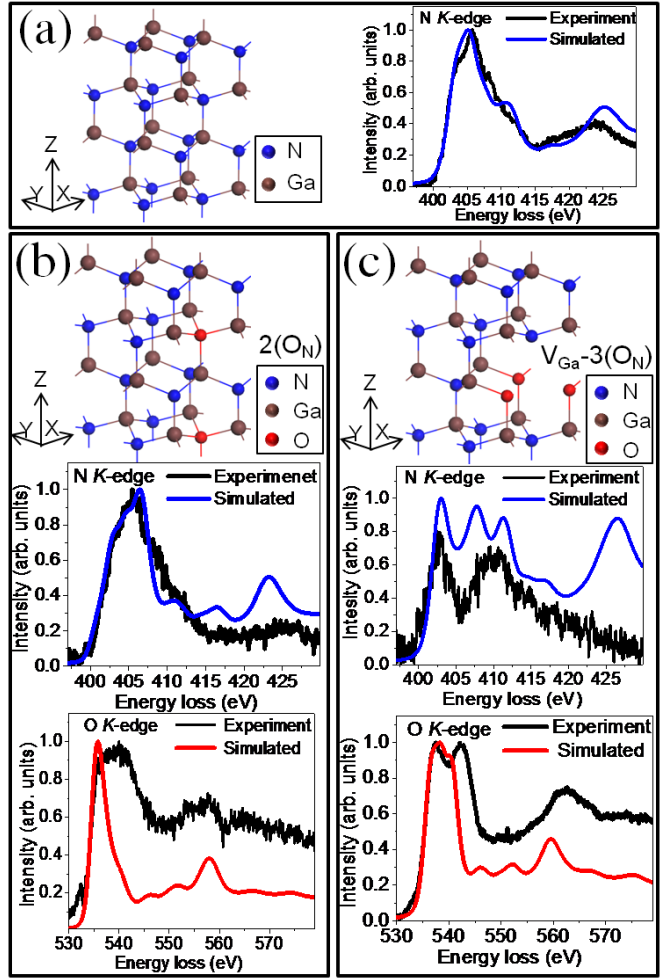

**Figure 2**. Experimental and simulated EELS spectra of GaN nanowires. (a) Right side; a simulated pure GaN structure, projected along Z-axis ∥ (0001) and Left side; the corresponding core loss N *K*-edge spectrum compared with that of the experimental spectrum collected from the nanowires of sample R4. (b) Top; A simulated GaN lattice with 2($O_N$) oxygen defect configuration, Middle and Bottom; simulated core loss N *K*-edge and O *K*-edge spectra, respectively, are compared with that of the experimental spectra collected from the nanowires of sample R3. (c) Top; A simulated GaN lattice with $V_{Ga}$–3($O_N$) oxygen defect configuration, Middle and Bottom; simulated core loss N K-edge and O K-edge spectra, respectively, are compared with that of the experimental spectra collected from the nanowires of sample R1.

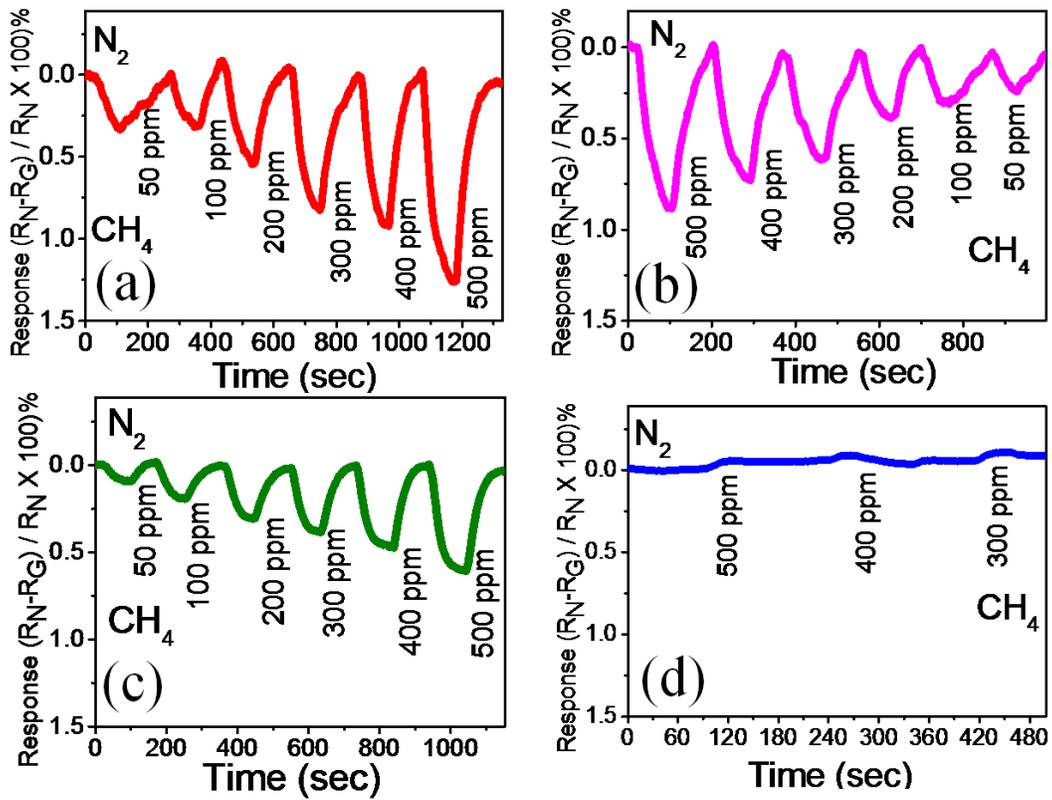

**Figure 3**. Temporal responses of CH$_4$ sensing by four devices made of nanowire samples (a) R1, (b) R2, (c) R3, and (d) R4 at different gas concentrations.

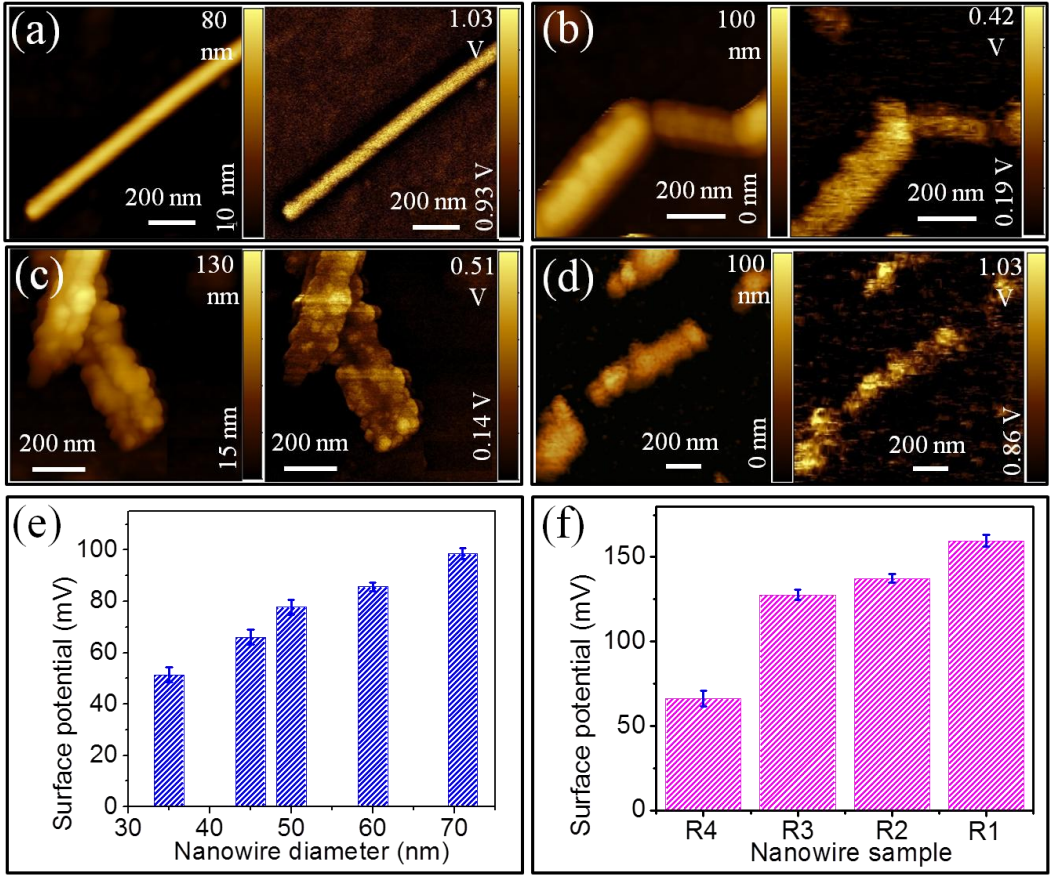

**Figure 4**. Typical SKPM micrographs of nanowire samples (a) R4, (b) R3, (c) R2, and (d) R1. In all images, Left side; topography maps and Right side; SP maps. (e) Nanowire diameter dependent SP values extracted from the SKPM maps of sample R4 and (f) SP plot of nanowires samples R1, R2, R3, and R4 with different oxygen impurity.

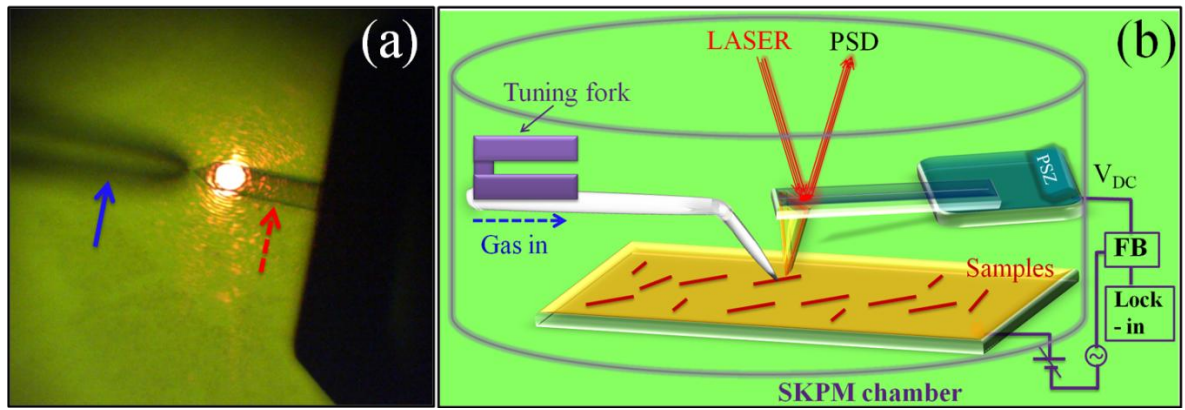

**Figure 5**. (a) Optical image of the aligned two probes used for the in situ SKPM measurements under gas exposure. Cantilever based Au coated Si probe (dashed arrow line) with beam-bounce feedback is used for SKPM measurements on nanowire samples dispersed on Au film. Non-optical normal-force feedback is used for the tuning fork based nano-pipette gas delivery probe (solid arrow line). (b) Schematic view of the SKPM study with the in situ gas exposure multiprobe setup.

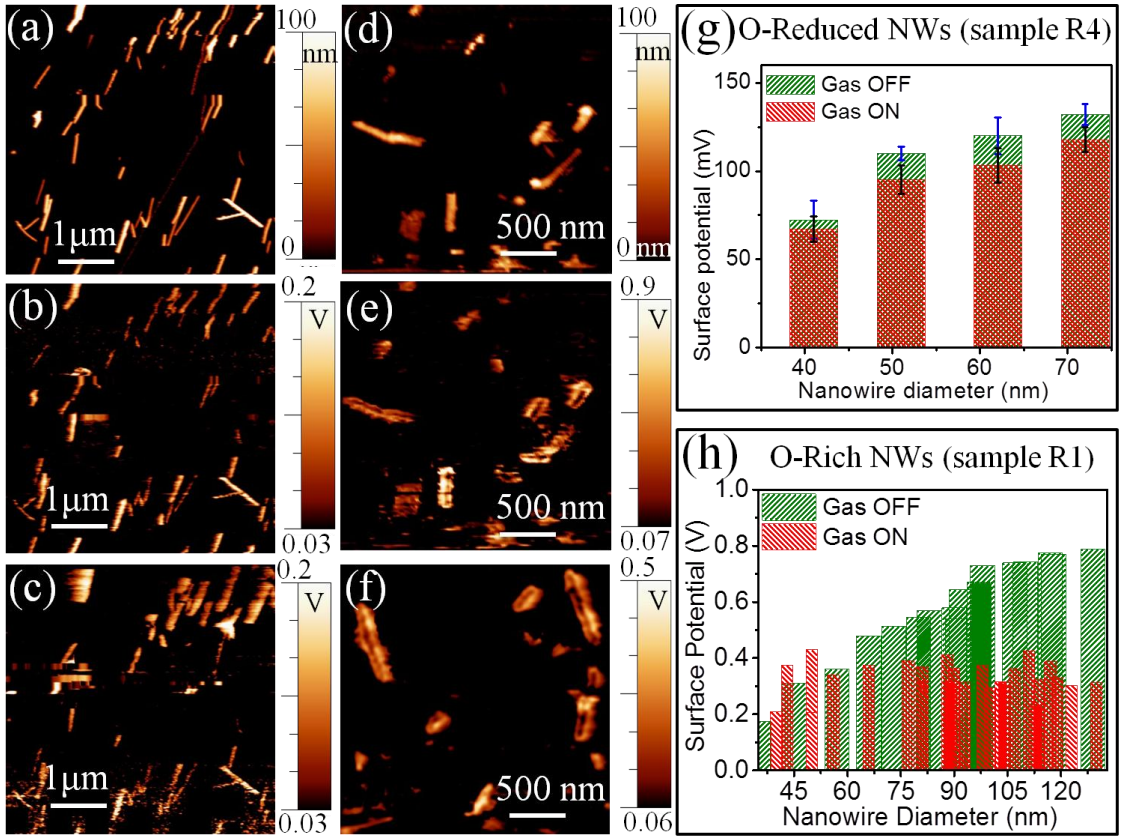

**Figure 6**. Typical SKPM micrographs of nanowires of sample R4 (Left side) and sample R1 (Right side) acquired at 100 °C and $10^{-2}$ mbar. Topography maps of samples (a) R4 and (d) R1. Typical SP maps of samples (b) R4 and (e) R1 before $CH_4$ exposure. Typical SP maps of samples (c) R4 and (f) R1 during the $CH_4$ exposure. (g) and (h) Nanowire diameter dependent SP values extracted from the SKPM maps of sample R4 and sample R1 respectively, before (green in colour) and during the $CH_4$ exposure.

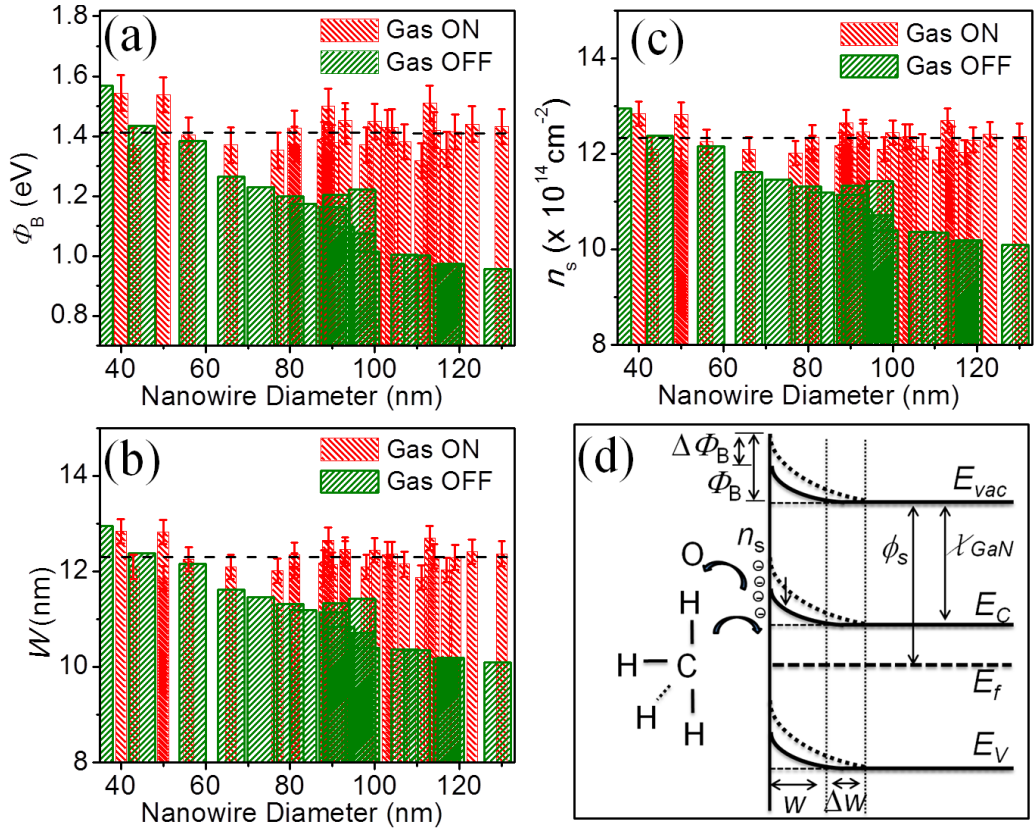

**Figure 7**. Nanowire diameter dependent (a) SBB ($\Phi_B$), (b) depletion width ($W$) and (c) surface charge density ($n_s$) values extracted from the SKPM maps of sample R1, before (Green) and during (Red) the $CH_4$ exposure. Horizontal dashed lines indicate the mean saturation values in all the plots. (d) Schematic view of the band diagram of GaN nanowires showing SBB ($\Phi_B$) and charge transfer process during $CH_4$ chemisorption.

**Supporting information:**

**Current–voltage (*I-V*) characteristics:**

Prior to the gas sensing measurements of the GaN nanowires, the nature of the electrical contacts to the nanowires was studied using *I-V* measurements. The *I-V* characteristics of the four devices, measured under -2 to +2 V bias, showed that the contacts are Ohmic in nature (Fig. S1). Resistances of the devices increased from 0.94 kΩ(sample R1) to 4.91 kΩ(samples R4) with a decrease in the oxygen concentration of the GaN nanowires. This observation indicates a strong influence of background electrons produced by oxygen shallow donors which act as *n*-type dopants in GaN.[1,2]

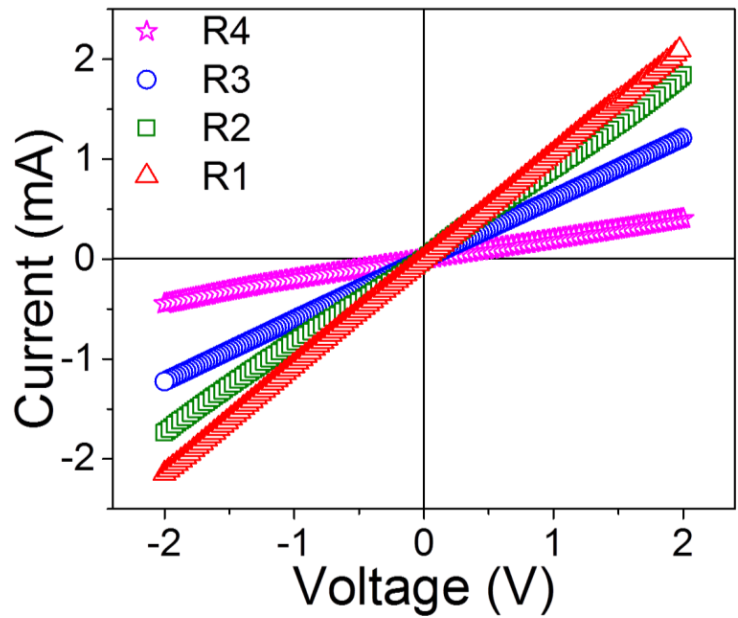

**Figure S1**. The current-voltage characteristics of the four devices made of samples R1, R2, R3, and R4 measured under ±2 V bias showing the contacts are Ohmic in nature and the increase in resistance for the samples R1 to R4.

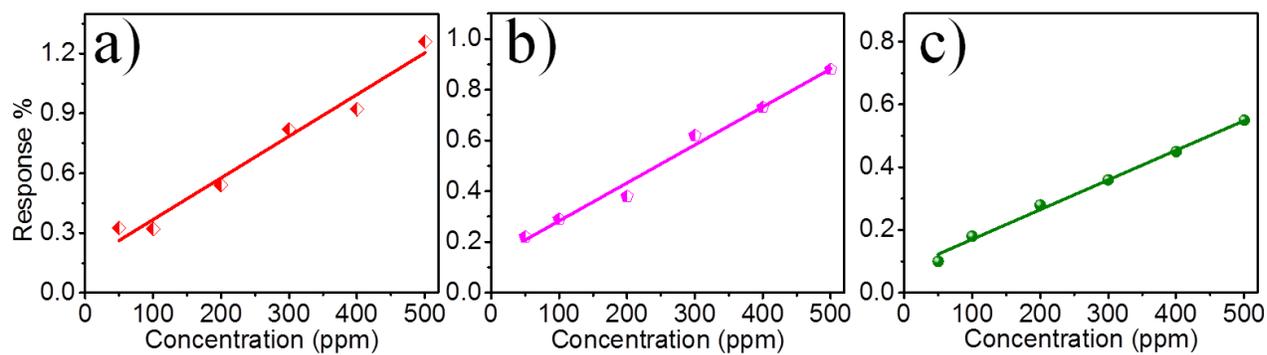

**Figure S2**. The sensor response over a concentration range of 50 – 500 ppm of $CH_4$ for the device (a) R1, (b) R2, and (c) R3.

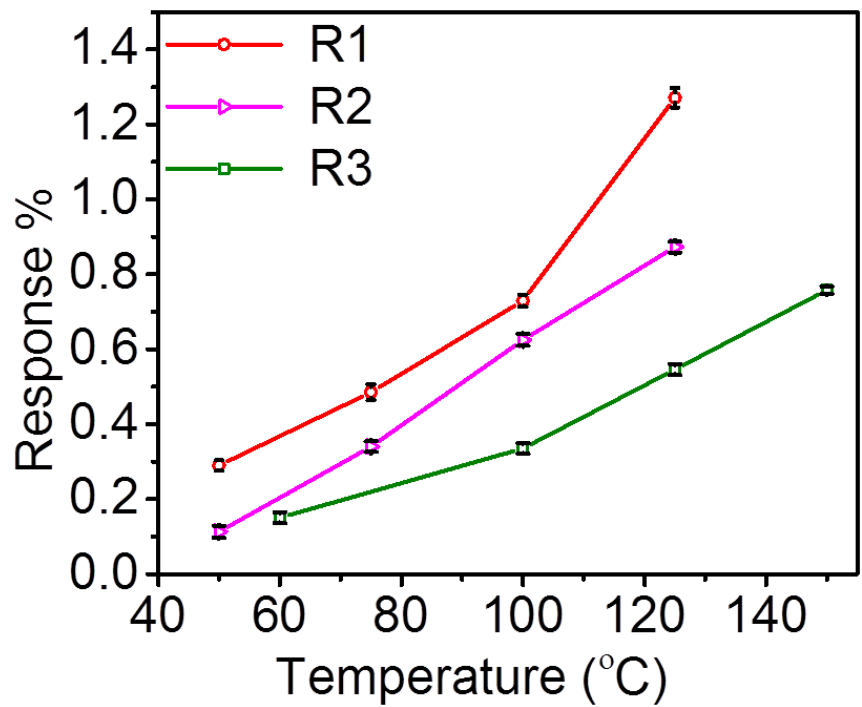

**Figure S3**. Temperature dependent $CH_4$ sensing response by the devices of samples R1, R2 and R3 showing the increase in response while increasing temperature.

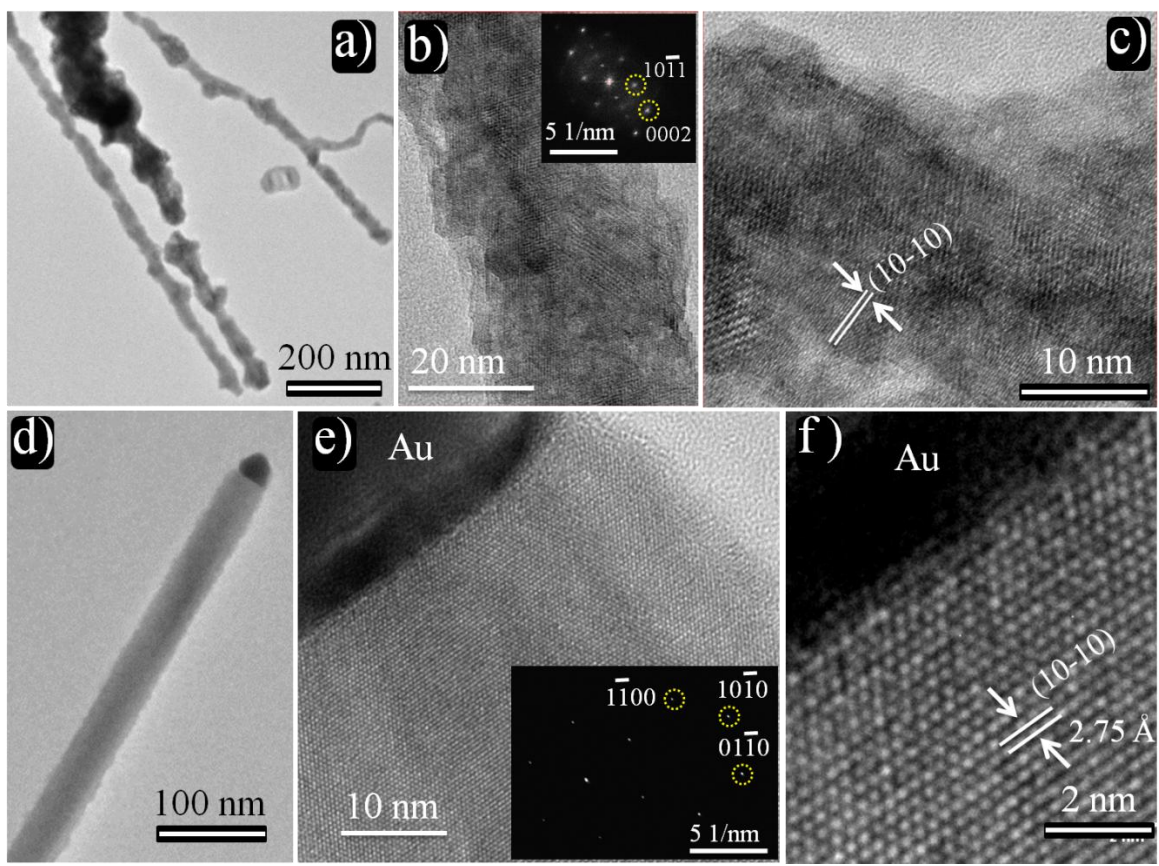

**Figure S4**. Structural studies of the nanowires: Sample R1; (a) Typical low magnification transmission electron microscopic (TEM) image of nanowires shows an irregular surface morphology, (b) high resolution TEM (HRTEM) image of the nanowire with non-uniform contrast along the wire. Inset shows fast-Fourier transform (FFT) image corresponding to HRTEM micrograph of the nanowire. (c) Magnified few of the HRTEM image of the nanowire with lattice fringes corresponding to {10$\bar{1}$0} planes of GaN phase. Sample R4; d) Low magnification TEM image of a typical nanowire, (e) HRTEM image collected near the tip of the nanowire having Au nanoparticle. Inset shows selected area electron diffraction (SAED) pattern of the nanowire, indexed to wurtzite GaN with zone axis along [0001] (f) Magnified view of HRTEM image shows the nanowire having nonpolar planes (10$\bar{1}$0).

**Structural Studies:**

Typical TEM micrograph of nanowires from the sample R1 shows an irregular surface morphology (Fig. S4(a)). The diffraction spots enclosed by doted circles in FFT image (inset of Fig. S4(b)) corresponding to HRTEM micrograph of the nanowire, are indexed to {10-10} and {0002} planes of wurtzite GaN phase. High resolution image of the nanowire shows non-uniform contrast (Fig. S4(b)) which may originate from variation in thickness along the nanowire and barely seen for lattice fringes. Lattice fringes with *d*-spacing of 0.275 nm corresponding to {10$\bar{1}$0} planes of GaN phase are observed in the magnified view (Fig. S4(c)) of high resolution image. A typical nanowire of the sample R4 shows a perfect rod like shape with uniform surface morphology (Fig. S4(d)) and having an Au catalyst particle on the tip. Uniform contrast along with sharp interface between GaN nanowire and Au catalyst particle is observed in the HRTEM image (Fig. S4(e)). The SAED pattern (inset of Fig. S4(e)) reveals that the nanowire is a single crystalline wurtzite phase of GaN with zone axes [0001]. An interplanar spacing of 0.275 nm, shown in the magnified view (Fig. S4(f)), corresponds to the nonpolar {10$\bar{1}$0} planes of wurtzite GaN. The growth direction of the nanowire is found to be along [10$\bar{1}$0] direction.

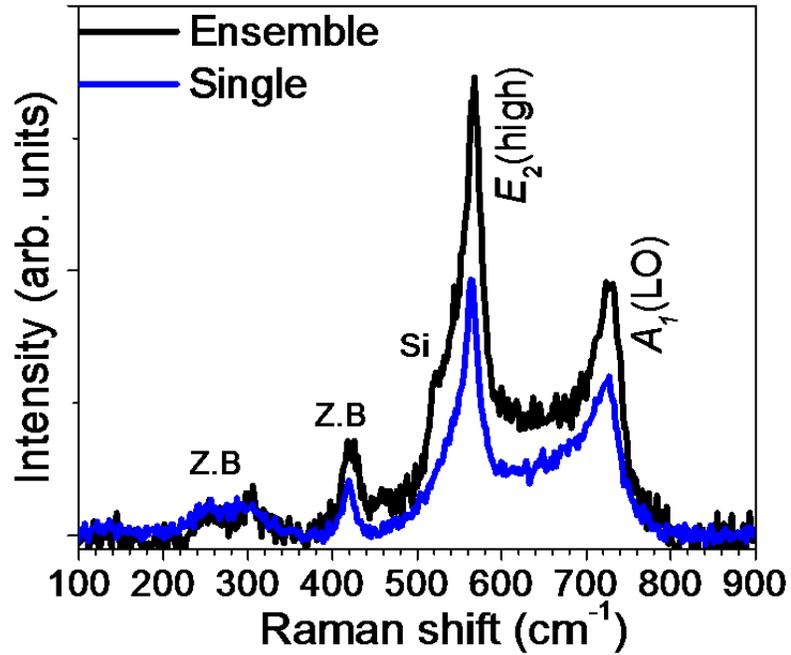

**Figure S5**. Raman spectra of ensemble GaN nanowires grown on Si substrate and a single GaN nanowire (laid on Al substrate to avoid Si background signal), of sample R1 grown under oxygen rich condition. No Raman modes corresponding to the oxide phase was observed either in single or ensemble GaN nanowires.

**SKPM tip calibration:**

Before proceeding to the SKPM measurements on the nanowire samples, Au coated Si tip was calibrated on pure Au film under similar conditions of the SKPM measurements. An absolute concentration of 1000 ppm $CH_4$ (99.999%) in $N_2$ (99.999 %) base gas was delivered through the nano-pipette probe which allowed the exposure of few ppm of sensing gas mixture with respect to the volume of the SKPM chamber. The very low volume of the gas flown in few ppm levels did not alter the base pressure of the chamber (~$10^{-2}$ mbar) over a period during the SKPM measurements. The RMS surface roughness value of ~1.5 nm for the Au film at 100 °C under $10^{-2}$ mbar (Fig. S6(a)) and the RMS value of CPD ($CPD_{RMS}$) ~ 28 mV (Fig. S6(b)) were measured without the $CH_4$ exposure. During the $CH_4$ exposure, the $CPD_{RMS}$ value (~ 25 mV) was not changed considering the estimated error limit (Fig. S6(c)). The results showed that the tip work function did not change with the experimental conditions in the SKPM chamber.

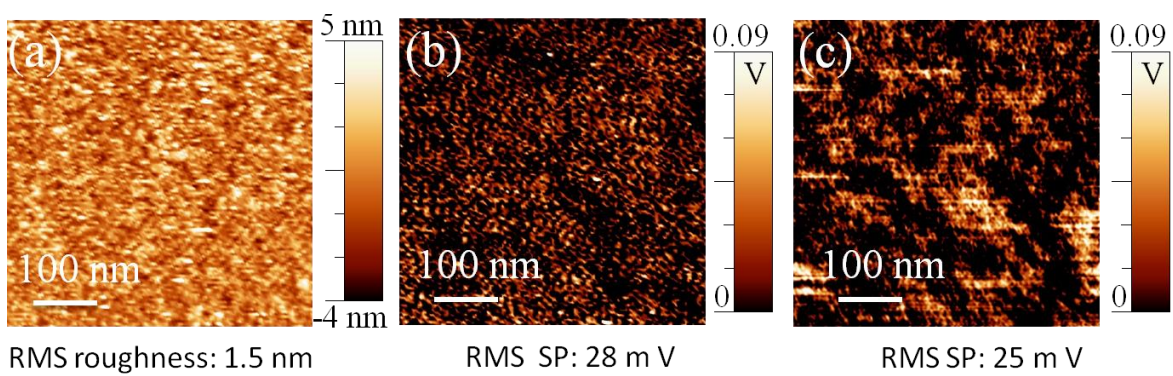

**Figure S6**. SKPM measurements on standard Au film for calibration of Au coated Si tip at 100 °C under $10^{-2}$ mbar. (a) Topography maps; the RMS roughness is ~1.5 nm. SP maps (b) before and (c) during the $CH_4$ exposure; RMS values ($CPD_{RMS}$) before and during the gas exposure are ~ 28 mV and ~25 mV, respectively.

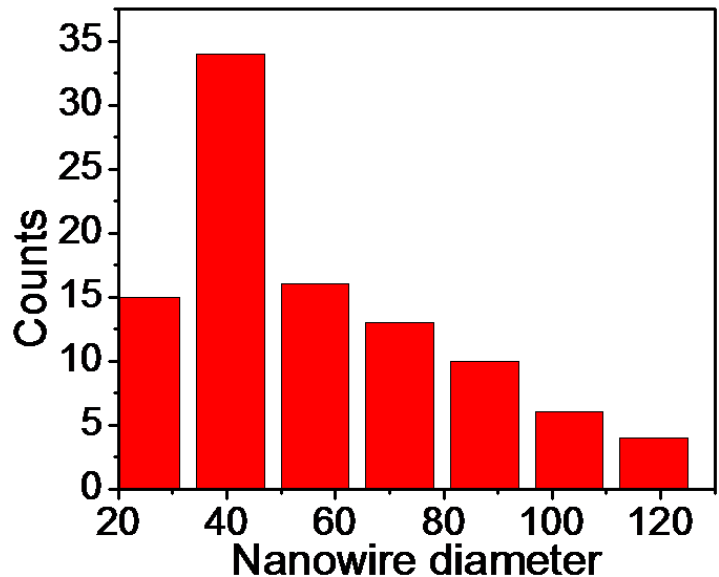

**Figure S7**. Diameter distribution of nanowires sample R1 grown under oxygen rich condition. A large number of nanowires having the diameter less than 60 nm are present.

**References:**

1. Reshchikov, M. A.; Morkoç, H. Luminescence properties of defects in GaN. *J. Appl. Phys.* **2005**, *97*, 061301.

2. Ptak, A.; Holbert, L.; Ting, L.; Swartz, C.; Moldovan, M.; Giles, N.; Myers, T.; Van Lierde, P.; Tian, C.; Hockett, R. Controlled oxygen doping of GaN using plasma assisted molecular-beam epitaxy. *Appl. Phys. Lett.* **2001**, *79*, 2740-2742.